\documentstyle[12pt]{article}

\newcommand{\beq}{\begin{equation}}
\newcommand{\eeq}{\end{equation}}

\begin{document}
\def\lag{\langle}
\def\rag{\rangle}
\vspace{2.5cm}
\title{A New Approach to Spin Glass Simulations\footnote{Submitted 
to Physical Review Letters.}}

\author{Bernd A. Berg$^{1,2}$ and Tarik Celik$^{1,3}$\\[3ex]
$^1$ Supercomputer Computations Research Institute\\
                      Tallahassee, FL~32306, USA\\[2ex]
$^2$ Department of Physics, The Florida State University\\
                      Tallahassee, FL~32306, USA\\[2ex]
$^3$ On leave of absence from Department of Physics\\
 Hacettepe University\\
 Ankara, Turkey}

\maketitle
\newpage
\begin{abstract}

We present a recursive procedure to calculate the parameters of the 
recently introduced multicanonical ensemble and explore the approach
for spin glasses. Temperature dependence of the energy, the entropy 
and other physical quantities are easily calculable and we report results 
for the zero temperature limit. Our data provide evidence that the large 
$L$ increase of the ergodicity time is greatly improved. The
multicanonical ensemble seems to open new horizons for simulations of
spin glasses and other systems which have to cope with conflicting 
constraints.
\end{abstract}

\newpage

The theoretical understanding of spin glasses, for a review see 
\cite{Bi1}, has remained a great challenge. In particular the low
temperature limit leaves many open questions about the effects of disorder
and frustration. For instance, it has remained controversial whether
Parisi's \cite{Paris} mean field theory provides the appropriate 
description for 3D spin glasses. The attractive alternative is the droplet
model \cite{Drop}, which in turn is equivalent to a one parameter scaling
picture \cite{Bray}. The simplest spin glass system to study such questions
numerically is the Edwards-Anderson model. In its Ising version it is 
described by the Hamiltonian
$$ H\ =\ - \sum_{<ij>} J_{ij} s_i s_j , \eqno(1) $$
where the sum goes over nearest neighbours and the exchange interactions
$J_{ij}=\pm 1$ between the spins $s_i=\pm 1$ are quenched random variables.
In our investigation we impose the constraint $\sum J_{ij} = 0$ for each
realization. Recent simulations \cite{Cara} of the 3D model in a magnetic 
field support the mean field picture. However, one may argue that 
sufficiently low temperatures on sufficiently large systems have not 
been reached. For previous simulations of the Edwards-Anderson model 
without magnetic field see \cite{Bhatt,SweWa}.

\begin{figure}
\vspace{10cm}
\caption[fig1]{Ising model magnetic probability density from $50\times
50$ lattice.} 
\end{figure}

Low temperature simulations of spin glasses suffer from a slowing down
due to energy barriers. To illustrate the problem, let us consider 
a simple ferromagnet: the 2D Ising model on an $50\times 50$ lattice.
In Figure~1 we give its magnetic probability density versus 
$\hat\beta = T^{-1}$. The two distinct branches below the Curie temperature
are associated with free energy valleys in configuration space, each of
which defines a (pure) thermodynamic state. At temperatures below the 
Curie point the ergodicity time$^1$
\footnotetext[1]{{It is for the present investigation of spin glasses 
more appropriate to use the term 
ergodicity time $\tau^e$, instead of tunneling time $\tau^t$ which is
appropriate in the context of surface tension investigations.}} 
$\tau_L^e$ increases exponentially fast with lattice size, asymptotically 
like $\exp [ f^s(\beta ) L^{D-1} ]$, where $f^s$ is the surface free
energy. Therefore, on large lattices at sufficiently low temperature the 
simulation of the system will, given a reasonable finite amount of 
computer time, never tunnel from one phase to the other. Besides for 
particular problems (like the determination of the order-order surface 
free energy) this lack of tunneling does not pose a major 
handicap for Ising model simulations. The reason is that the two 
configuration space valleys are related by the exact symmetry 
$s_i \to -s_i$ of the Hamiltonian. Exploring one valley by means of a 
simulation yields also all the properties of the other one and, hence, 
allows to overlook the entire system. 

The situation is much more involved for spin glasses. For low enough
temperature the system is supposed to split of into many thermodynamic
states, separated by similar tunneling barriers as the two pure states
of the Ising model. However, unlike in the Ising model the states are not 
related to each other by a symmetry of the Hamiltonian. Rather they appear
because of accidental degeneracy which in turn occurs because of randomness
and frustration of the system. For computer simulations this means that 
one would like to explore many 
independent configuration space valleys while keeping track of their 
relative weights. The groundstate energies associated with these valleys 
may or may not be degenerate, but it should be noted that even if they 
are not degenerate, tunneling between the valleys would still be governed
by the energy barriers. The physics of these barriers is far less well 
understood as in the ferromagnetic case. As detailed finite size scaling 
(FSS) studies do not exist, it is unclear to us to what extent these 
barriers depend on the system size, whereas the temperature dependence has 
been investigated \cite{Bi1}. We use the notation bifurcation temperature 
(bifurcation point) for the temperature at which the spin glass 
configuration space (phase transition or not) begins to split of into a 
number of valleys which are well separated by energy barriers. In the 
present paper we suggest that the increase of $\tau^e_L$ can be reduced 
to a fairly decent power law by performing a simulation which covers in a 
single ensemble a whole temperature range from well above to far below 
the bifurcation point. The appropriate formulation is provided by a
generalization of the multicanonical ensemble \cite{our1}.
To test the approach we have performed 
multicanonical simulations of the 2D Ising ferromagnet and then of
the 2D Edward-Anderson Ising spin glass model.
We concentrate on ground state properties
what is a kind of worst case scenario for the performance of the
multicanonical simulation. It should be noted that Figure~1 does 
rely on a multicanonical simulation, what 
explains the slight asymmetry between the two branches. 

Ever since the pioneering paper by Metropolis et al.\cite{Metro} most MC 
simulations concentrated on importance sampling for the canonical Gibbs 
ensemble. It has always been well-known, for instance \cite{Bi0},
that one is allowed to choose phase-space points according to any other
probability distribution, if it is convenient. However, a systematic 
reasoning for a better than canonical choice has rarely been put forward$^1$.
\footnotetext[1]{{Notable may be microcanonical simulations, but for 
the ergodicity problems on which we focus here they perform even worse 
than the canonical approach does.}}
It is our suggestion that in a large class of situations, 
in particular those where canonical simulations face severe ergodicity 
problems, it is more efficient to reconstruct the Gibbs 
ensemble from a simulation of a multicanonical ensemble \cite{our1}
than simulating it directly. In canonical simulations configurations
are weighted with the Boltzmann factor $P_B (E) = \exp (-{\hat\beta} E)$.
Here $E$ is the energy of the system under consideration.
The resulting canonical probability density is 
$$ P_c (E)\ \sim\ n(E) P_B(E), \eqno(2) $$
where $n(E)$ is the spectral density. In order of increasing severity
problems with canonical spin glass simulations are:

\noindent
{\it i}) Simulations at many temperatures are needed to get an overview 
         of the system.

\noindent
{\it ii}) The normalization in equation (2) is lost. It is 
          tedious to calculate important physical quantities like the 
          free energy and the entropy.

\noindent
{\it iii}) The low temperature  ergodicity time $\tau_L^e$ diverges 
           fast with lattice size (either exponentially or with a
           high power law). The relative weights of pure states
           can only be estimated for small systems.

Let us choose an energy range $E_{\min}\le E\le E_{\max}$ and
define for a given function $\beta (E)$ the function $\alpha (E)$ by 
the recursion relation (with the Hamiltonian~(1) the energy changes
in steps of 4) 
$$ \alpha (E-4)\ =\ \alpha (E) + 
\left[ \beta (E-4) - \beta (E) \right] E,\ 
\alpha(E_{\max}) = 0  . \eqno(3) $$
The multicanonical ensemble \cite{our1} is then defined by weight factors
$$P_M (E)\ =\ \exp \left[ -\beta (E) E + \alpha (E) \right] ,\eqno(4)$$
where $\beta (E)$ is determined 
such that for the chosen energy range the resulting multicanonical 
probability density is approximately flat:
$$ P_{mu} (E)\ =\ c_{mu}\ n(E) P_M (E)\ \approx\ {\rm const.} \eqno(5) $$
In the present study we take $E_{\max} =0$ ($\beta (E) \equiv 0$ for
$E\ge E_{\max}$) and $E_{\min}=E^0$ the ground state energy of the 
considered spin glass realization. 

A multicanonical function $\beta (E)$ can be obtained  via 
recursive MC calculations. One performs simulations $\beta^n (E)$, 
$n=0,1,2,...$, which yield probability 
densities $P^n (E)$ with medians $E^n_{\rm median}$. For 
$E < E^n_{\min} < E^n_{\rm median}$ the probability density $P^n (E)$ 
becomes unreliable due to insufficient statistics, caused by the  
exponentially fast fall-off for decreasing $E$. We start off with $n=0$
and $\beta^0 (E) \equiv 0$. The recursion from $n$ to $n+1$ reads  
$$ \beta^{n+1} (E)\ =\ 
\cases{ \beta^n (E)\ {\rm for}\ E \ge E^n_{\rm median}; \cr
\beta^n (E^n_{\rm median})\ + \cr
(E^n_{\rm median}-E)^{-1}\ \ln \left[ P^n(E^n_{\rm median})/P^n(E) \right]\
                         {\rm for}\ E^n_{\rm median} > E \ge E^n_{\min}; \cr
\beta^{n+1} (E^n_{\min})\ {\rm for}\ E < E^n_{\min} . \cr} \eqno(6) $$
Here the $n^{\rm th}$ simulation may be constrained to 
$E < E^{n-1}_{\rm median}$ by rejecting all proposal with energy 
$E > E^{n-1}_{\rm median}$, but one has to be careful with such bounds
in order to maintain ergodicity. The recursion is stopped for $m$ with 
$E^{m-1}_{\min}=E^0$ being groundstate.
Starting with this simple approach we have explored several more 
sophisticated variants. Considerable speed-ups and gains in stability could 
be achieved. The CPU time spent to estimate the multicanonical parameters
was 10\% to 30\% of the CPU time spent for simulations with the final set. 

Once the functions $\beta (E)$ and $\alpha (E)$ are fixed, the multicanonical
simulation exhibits a number of desireable features:

\noindent {\it i}) By reweighting with 
$\exp [-\hat\beta E + \beta (E) E - \alpha (E)]$ the canonical expectation
values
$$ {\cal O} (\hat\beta)\ =\ Z({\hat\beta} )^{-1}
\sum_E {\cal O} (E)\ n(E)\ \exp (-\hat\beta E) , \eqno(7) $$
where $Z(\hat\beta ) = \sum_E n(E)\ \exp (-\hat\beta E)$ is the partition 
function, can be reconstructed for all $\hat\beta$ in an entire 
range $\beta_{\min} \le \hat\beta \le \beta_{\max}$. Here 
$\beta_{\min} = \beta (E_{\max})$ and $\beta_{\max} = \beta (E_{\min})$ 
follow from the requirement 
$E_{\max} \ge E (\hat\beta ) \ge E_{\min}$, and $E(\hat\beta )$ 
follows from (7) with ${\cal O} (E) = E$. With our choice $E_{\max}=0$ 
and $E_{\min}=E^0$ groundstate, $\beta_{\min}=0$ and 
$\beta_{\max}=\infty $ follows.

\noindent {\it ii}) The normalization constant $c_{mu}$ in equation (5) 
follows from $Z(0)=\sum_E n(E) = 2^N$, where $N$ is the total number of
spin variables. This gives the spectral density and allows to calculate 
the free energy as well as the entropy. 

\noindent {\it iii}) We conjecture that the slowing down of canonical 
low temperature spin glass simulations becomes greatly reduced.
For the multicanonical ensemble it can be argued \cite{our1}
that single spin updates cause a 1D random walk behaviour of the energy $E$.
As $E_{\max} - E_{\min} \sim V$, one needs $V^2$ updating steps to cover
the entire ensemble. For first order phase transition the observed slowing
down was only slightly worse than this optimal behaviour.
Our present MC data show more drastic modifications for spin glass 
simulations.

To quantify our discussion of the slowing down, we have to define the
ergodicity time $\tau^e_L$. Roughly speaking it is the CPU time needed
to collect independent configurations of the system, for instance 
groundstates when this is the main interest. Regarding the definition
of $\tau^e_L$, an $L$-independent over-all factor is free.
We define $\tau^e_L$ as the average number 
of sweeps needed to move the energy from $E_{\max}$ to $E_{\min}$ and
back. A sweep is defined by updating each spin on the lattice once
(in the average). It should
be noted that the number of ``tunneling'' events with respect to the
ergodicity time gives a lower bound on the number 
of independent groundstates samples.
This follows from another trivial, but remarkable property of the 
multicanonical ensemble: Each time a sweep is spent at $\beta (E)\equiv 0$,
the memory of the previous Markov chain is lost entirely, and a truly 
independent new series of configurations follows. 
The condition $E \ge E_{\max}$ is appropriate to substitute for the 
somewhat too strict constraint of a sweep at $\beta\equiv 0$.
As a corollary: with a disordered starting 
configuration the multicanonical ensemble is immediately in equilibrium.

As an exercise and to check our code on exact results, we performed a 
multicanonical simulation of the 2D Ising model with 
$0\le\hat\beta < \infty$. We kept the time series of
two million sweeps and measurements on a $25\times 25$ lattice and
verified that the finite lattice 
specific heat results of Ferdinand and Fisher \cite{Ferd}
are well reproduced. No difficulties are encountered with the
multicanonical ensemble when crossing the phase transition point. 
To explore the possibility of zero temperature entropy
calculations, we used $Z(0)=2^{625}$ as input and obtained 
$S^0 = 0.61 \pm 0.09$ for the total groundstate entropy. This 
corresponds to an estimate of $1.84 \pm 0.17$ groundstates, {\it i.e.}
within statistical errors we are in agreement with two. Using
$Z(0)=2^{2500}$ and a time series of four million sweeps on a $50\times 50$
lattice we obtained $2.07 \pm 0.22$ for the number of groundstates. 
Figure 1 is obtained from the simulation of this lattice.

After this test we turned to the 2D Edwards-Anderson spin
glass. On lattices of size $L=4$, 12, 24 and 48 we performed multicanonical 
simulations. Up to $L=24$ we investigated ten different realizations per 
lattice and, due to CPU time constraints, we considered only five 
realizations for the $L=48$ lattice. The 
multicanonical energy distribution for one of our $L=48$ realizations 
is depicted in Figure~2. The fall-off for $-e < 0$ is like that of the 
canonical distribution at $\hat\beta =0$. For $0\le -e < -e^0$ an
impressive flatness (about 800 energy entries on the lattice under 
consideration) is quickly achieved by the recursion (6). Close to the
groundstate some fluctuations are encountered on which we comment 
elsewhere \cite{our3}. As it is not obvious from the scale of the 
figure, we like to remark that the groundstate is not the state with
the lowest number of entries, but a state close to it. Further,
it should be understood that deviations from the desired constant
behaviour (5) do only influence the statistical error bars, but not the
estimates themselves. Therefore, such deviations do not pose problems as 
long as they can be kept within reasonable limits of approximately one order 
of magnitude.

\begin{figure}
\vspace{10cm}
\caption[fig2]{Multicanonical energy density distribution for one
$L=48$ realization.} 
\end{figure}

\begin{figure}
\vspace{10cm}
\caption[fig3]{Ergodicity times versus lattice size on a double log scale.}
\end{figure}

The Table gives an overview of some of our numerical results. The 
ergodicity time $\tau^e_L$ is as defined above, and for 
$\beta_{\max}$ we take $\beta (E^0)$, where it should be noted that due to 
our computational procedure $\beta (E)$ is a noisy function. The reported
values and their error bars are obtained by combining the results 
from the different realizations, which enter with equal weights. 
In Figure 3 we plot the ergodicity time versus lattice size $L$ on
a log--log scale. The data are consistent with a straight line fit
($Q$ denotes the goodness of fit), which gives the
finite size behaviour
$$ \tau^e_L\ \sim\ L^{4.4 (3)} ~~~{\rm sweeps}. \eqno(8)$$
In CPU time this corresponds to a slowing down $\sim V^{3.2 (2)}$. It should
be remarked that a fit of form $\tau^e_L \sim \exp (c L)$ results
in a completely unacceptable goodness of fit $Q < 10^{-6}$. Still,
the behaviour (8) is by an extra volume factor worse than the close to 
optimal performance we hoped for. The reasons will be considered 
elsewhere \cite{our3}.

\begin{table}[h]
\centering
\begin{tabular}{||c|c|c|c||}                    \hline
$L$ & Statistics            & $\tau^e_L$            & $\beta_{\max}$ \\ \hline
 4  & $10\times 1,600,000$ &      $35.3\pm     2.8$ & $0.75\pm 0.06$ \\ \hline
12  & $10\times   760,000$*&     $2,607\pm     450$ & $1.47\pm 0.04$ \\ \hline
24  & $10\times 1,600,000$ &   $193,750\pm  43,820$ & $2.12\pm 0.11$ \\ \hline
48  & $~~5\times 5,200,000$*&$1,457,315\pm 516,925$ & $2.22\pm 0.07$ \\ \hline
\end{tabular}
\caption[tab1]{{\em Overview of some results. For the data points marked with *
the statistics for different realizations varies somewhat and average values
are given. }}
\end{table}

\begin{figure}
\vspace{10cm}
\caption[fig4]{FSS estimate of the infinite volume entropy per spin.}
\end{figure}

We estimate the infinite volume groundstate energy and entropy from
FSS fits of the form $f_L = f_{\infty} + c/V$. The entropy fit is 
depicted in Figure~4, and the energy fit looks similar. Our 
energy estimate is $e^0 = -1.394 \pm 0.007$, consistent with the 
previous MC estimate \cite{SweWa} $e^0 = -1.407 \pm 0.008$ as well as with 
the transfer matrix result \cite{CMc} $e^0 = -1.4024 \pm 0.0012$. Our
entropy estimate $s^0 = 0.081 \pm 0.004$ is also consistent
with the MC estimate \cite{SweWa} $s^0 = 0.071 \pm 0.007$, but barely
consistent with the more accurate transfer matrix result 
\cite{CMc} $s^0 = 0.0701 \pm 0.005$. 
Still, a larger statistical sample and a careful study of
systematic error sources would be needed to claim that there is a
significant discrepancy. Our results could be improved by exploiting 
the high temperature expansion of the entropy, as it was done in 
\cite{SweWa}. However, this would be against the spirit of this paper 
which is to explore the possibilities and limits of multicanonical spin 
glass simulations. 

It is not entirely straightforward to compare multicanonical and 
standard simulations. For instance autocorrelation times of 
multicanonical simulations come out short due to the triviality that the 
simulation spends most of its time at rather small effective $\beta $ 
values. Therefore, autocorrelations are not well suited. Our ergodicity
time, the average number of sweeps to find truly independent groundstates,
is a more useful quantity. When relying on it, {\it i.e} concentrating on
groundstate properties, we should remember that this pushes the 
multicanonical simulation to its extreme. This way of calculating
groundstate properties gives for free all properties in-between, from
$\hat\beta =0$ on. If a phase transition occurs, its study is also included 
(we noticed no particular difficulties in the 2D Ising model). For 
instance in the 3D Ising spin glass the canonical slowing down at 
$\hat\beta_c$ \cite{Bi1} is already worse than the one given by our 
equation (8).

Although the slowing down (8) is severe, it seems to provide an important 
improvement when compared with the slowing down which canonical 
simulations encounter for temperatures below the bifurcation temperature. 
For $L\ge24$ canonical simulations \cite{Bi1,Bhatt} are unable to 
equilibrate the systems at the $\beta_{\max}$ values reported in our 
table since the relaxation time is by far too long. Presumably due to
this fact the literature focused on other questions and does not provide
detailed FSS investigation of relaxation times. At the present stage we 
did not want to spent CPU time on extensive canonical simulations. 
Therefore, detailed comparisons are not possible. However, a rough 
estimate of the canonical ergodicity time may be \cite{Bi1} 
$ \tau^e_{\rm canonical}\ \sim\ \exp \left( C \hat\beta - C' \right)
~{\rm with}~~ C\approx 15 ~~{\rm and}~~ \ C' \approx 11.6 .$
The scaling with $L$ may then be hidden in the $L$ 
dependence of our $\beta_{\max}$ values, which is argued to be divergent
like $\beta_{\max} \sim \ln (V)$, and this line of reasoning gives a
slowing down of the canonical algorithm like $V^{15}$. With
$\beta_{\max}=2.12$ (our $L=24$ case) one gets a canonical ergodicity time 
of order $10^8-10^9$ when the missing constant is assumed to be of 
order one, whereas our $\tau^e_{24}$ is about $2\times 10^5$. Already for 
moderately large $\hat\beta$-values, a doubtful procedure has emerged in 
the literature \cite{Bhatt,Cara}:
Realizations which (according to a well defined criterium) cannot be
equilibrated are simply omitted. As these realizations are as legitimate
members of the statistical ensemble as any other choice of the quenched
random variables $J_{ij}$, an uncontrollable bias is inevitably introduced.
In contrast to this, we were able to equilibrate every single realization
encountered. 

When one is only interested in groundstate properties, minimization 
algorithms have to be considered. As a method simulated annealing 
\cite{Kirk} stands out because of its generality, although there are
more efficient algorithms for special cases, which should be used
when appropriate. In simulated annealing the results depend on the 
cooling rate $r=-\triangle T / {\rm sweeps}$. For our model the behaviour
$e (r) = e^0 + c~r^{1\over 4}$ with $c \approx 0.5$
 ($\triangle T = -0.1$ fixed) is indicated \cite{Grest}. 
To find a true groundstate, one has to reduce
$[e(r) - e_0]$ to the order $1/V$. Assuming that the constant $c$ 
is volume independent (only the lattice size $100\times 100$
was considered in \cite{Grest}), this translates to 
\# $ {\rm sweeps}\ \sim\ V^4 = L^8 $,
far worse than our equation (8). This result is kind of amazing as 
the multicanonical ensemble has eliminated directed cooling and is
nevertheless more efficient. If one does not insist on true
groundstates one can then relax the condition $[e(r) - e^0] \sim 1/V$.
For instance, any behaviour $[e(r) - e^0] \to 0$ with $L\to\infty$
would still give the correct density, and simulated annealing would
slow down less dramatic. On the other hand,
this would also imply a less stringent multicanonical simulation.

A comparison with the cluster-replica MC algorithm \cite{SweWa} is even less
clear cut. The obtained estimates of the groundstate energy and entropy are
in accuracy similar to ours. As one has to simulate many replica at
many $\hat\beta$-values a direct comparison is impossible. Clearly, the
results reported on slowing down are more promising than ours for the 
ergodicity time. On the other hand, to our 
knowledge the replica MC approach has never been applied to the 3D 
Edwards-Anderson model, and one may well encounter difficulties.
In contrast, 3D multicanonical simulations are straightforward. In fact 
the dimension is just a parameter in all our computer programs and we 
have already carried out various 3D test runs.
Let us address the question of algorithms from a more general perspective.
With a Metropolis type implementation our optimum performance will be
bounded from below by a slowing down $\sim V^2$ in CPU time,  
and we are outperformed
by any algorithm which can do better than this. For a number of important,
but often highly specialized, applications such better algorithms exist
and should be used. The main advantage of 
the multicanonical ensemble is its generality. With this respect our
method resembles simulated annealing \cite{Kirk}, while clearly avoiding 
some of its disadvantages. Most notably, the 
relationship to the equilibrium canonical ensemble remains exactly 
controlled. Finally, it should be stressed that the multicanonical
ensemble is an ensemble and not an algorithm. One may find better algorithms
than conventional Metropolis updating to simulate this new ensemble. To try 
a combination with cluster algorithms is an attractive idea.

Our results make clear that the 
multicanonical approach is certainly a relevant enrichment of
the options one has with respect to spin glass simulations. 
The similarities of spin glasses to other problems with 
conflicting constraints \cite{Kirk} suggest that multicanonical 
simulations may be of value for a wide range of investigations:
optimization problems like the travelling salesman, neural networks, 
protein folding, and others.
Multicanonical simulations of the 3D Edwards-Anderson model may eventually 
shed new light on the question whether the model exhibits mean field like 
behaviour or some kind of droplet picture applies \cite{Paris,Drop,Bray,Cara}.
It seems that previous numerical work remained somewhat inconclusive as 
sufficiently low temperatures could not be reached without destroying 
the thermodynamic equilibrium.

\section*{Acknowledgements}

We would like to thank Ulrich Hansmann and Thomas Neuhaus for numerous
useful discussions. Further, we are indebted to Philippe Backouche,
Norbert Schultka, and Unix\_Guys for valuable help.
T.C. was supported by TUBITAK of Turkey and likes to thank Joe Lannutti
and SCRI for the warm hospitality extended to him.
Our simulations were performed on the SCRI cluster of fast RISC
workstations. 

This research project was partially funded by the the 
Department of Energy under contracts DE-FG05-87ER40319, 
DE-FC05-85ER2500, and by the NATO Science Program.

\bibliographystyle{unsrt}

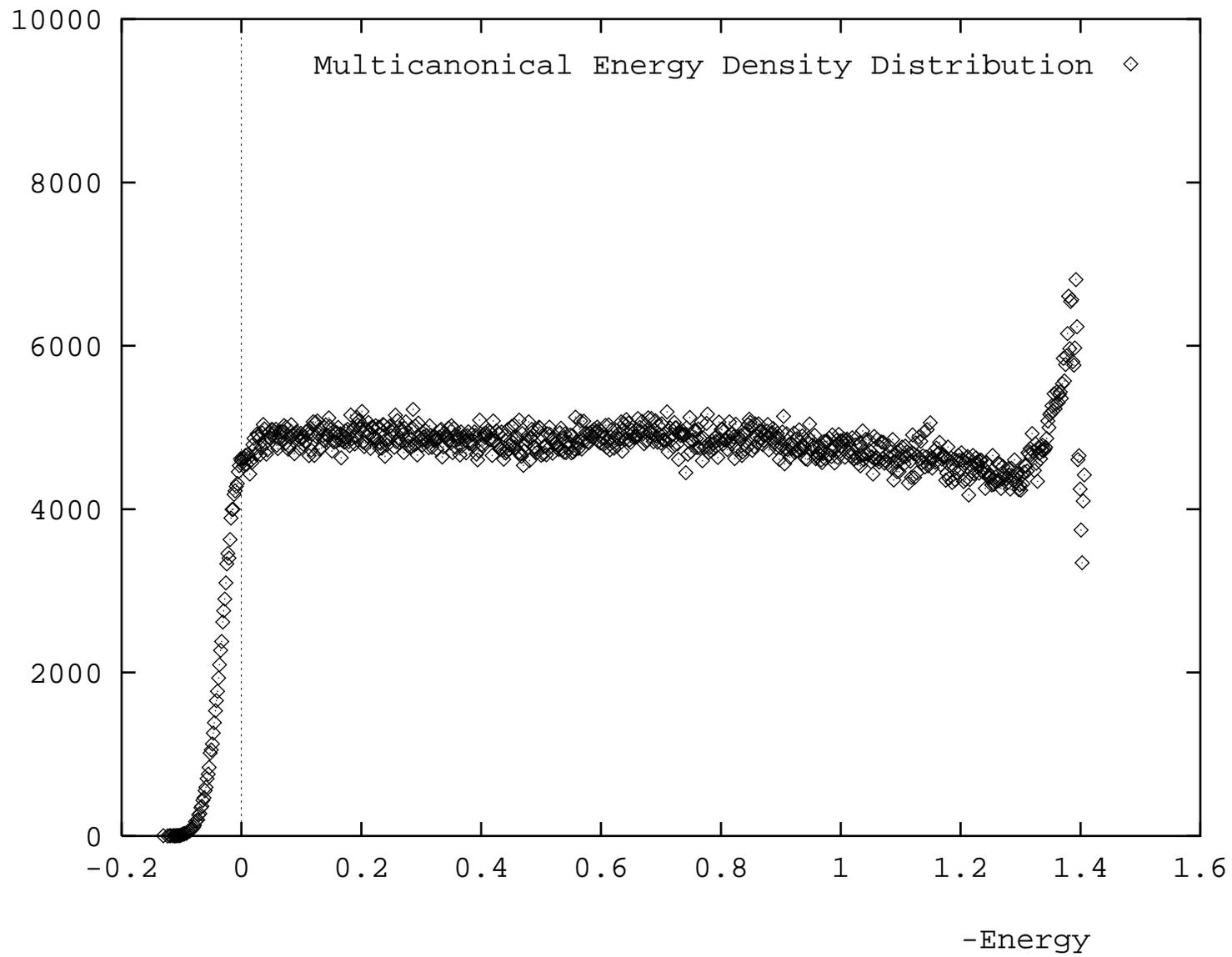
\end{document}